\documentclass[twocolumn]{revtex4}
\usepackage{epsfig}
\usepackage{amssymb}
\usepackage{amsmath}
\usepackage{amsfonts}

\newcommand{\R}{\mathbb{R}}

\newcommand{\fm}{\mathfrak{m}}

\newcommand{\be}{\begin{equation}}
\newcommand{\ee}{\end{equation}}
\newcommand{\bea}{\begin{eqnarray}}
\newcommand{\eea}{\end{eqnarray}}
\newcommand{\nn}{\nonumber}
\newcommand{\kt}{\rangle}
\newcommand{\br}{\langle}

\newcommand{\ed}{\end{document}}

\newcommand{\bi}{\begin{itemize}}
\newcommand{\ei}{\end{itemize}}

\begin{document}

\title{Path-Integral Formulation of Pseudo-Hermitian
Quantum Mechanics and\\ the Role of the Metric Operator}

\author{A.~Mostafazadeh}
\address{Department of Mathematics, Ko\c{c} University, Sariyer 34450,
Istanbul, Turkey}

\begin{abstract}
We provide a careful analysis of the generating functional in the
path integral formulation of pseudo-Hermitian and in particular
${\cal PT}$-symmetric quantum mechanics and show how the metric
operator enters the expression for the generating functional.

\medskip

\hspace{6.0cm}{Pacs numbers: 03.65.Ca, 03.65.Db, 11.10.-z,
11.30.Er}
\end{abstract}

\maketitle

In the pseudo-Hermitian representation of quantum mechanics
\cite{jpa-2004b}, a quantum system is determined by a triplet
$({\cal H},H,\eta_+)$ where ${\cal H}$ is an auxiliary Hilbert
space, $H:{\cal H}\to{\cal H}$ is a linear (Hamiltonian) operator
with a real spectrum and a complete set of eigenvectors, and
$\eta_+:{\cal H}\to{\cal H}$ is a linear, positive-definite,
invertible (metric) operator fulfilling the pseudo-Hermiticity
condition \cite{p1}
    \be
    H^\dagger=\eta_+H\eta_+^{-1}.
    \label{ph}
    \ee
The physical Hilbert space ${\cal H}_{\rm phys}$ of the system is
defined as the complete extension of the span of the eigenvectors
of ${\cal H}$ endowed with the inner product
    \be
    \br\cdot,\cdot\kt_+:=\br\cdot|\eta_+\cdot\kt,
    \label{inn}
    \ee
where $\br\cdot|\cdot\kt$ is the defining inner product of ${\cal
H}$, and the observables are identified with the self-adjoint
operators acting in ${\cal H}_{\rm phys}$, alternatively
$\eta_+$-pseudo-Hermitian operators acting in ${\cal H}$,
\cite{jpa-2004b}.

Pseudo-Hermitian quantum mechanics \cite{jpa-2004b} has primarily
developed in an attempt to unravel the mathematical structures
responsible for the intriguing spectral properties of ${\cal
PT}$-symmetric Hamiltonians \cite{bender-prl-1998} such as
    \be
    H=\frac{p^2}{2m}+\frac{\mu^2 x^2}{2}+i\epsilon x^3,~~~~
    m\in\R^+,~~~\mu,\epsilon\in\R.
    \label{cubic-osc}
    \ee
These properties were originally noticed in dealing with the field
theoretical analogues of (\ref{cubic-osc}). It has also been
conjectured that such models may have important applications in
particle physics \cite{bender-ajp}. This has led a number of
researchers to investigate various ${\cal PT}$-symmetric field
theories \cite{PT-field}. Similarly to standard quantum field
theory, the path-integral techniques are indispensable in dealing
with ${\cal PT}$-symmetric field theories. Yet a careful analysis
of the path-integral formulation of Pseudo-Hermitian quantum
mechanics has been lacking till recently. The first careful
analysis of this problem is given in Ref.~\cite{jones-rivers}
where the authors use the Hermitian representation of the
corresponding pseudo-Hermitian models to determine the appropriate
source term in the generating functional $Z[J]$. The results of
\cite{jones-rivers} give the impression that the metric operator,
which is the central ingredient in the operator formulation of
pseudo-Hermitian quantum mechanics, does not play an important
role in the path-integral formulation of the theory. The purpose
of the present paper is to elucidate the role of the metric
operator in the latter formulation. This is indeed necessary for a
correct formulation of ${\cal PT}$-symmetric field theories. It
also provides a straightforward interpretation of the results
pertaining the metric-independence of thermodynamical quantities
associated with non-interacting pseudo-Hermitian statistical
mechanical systems that is discussed in \cite{jacobsky}.

Consider the definition of the generating functional (partition
function) that is the starting point of path-integral formulation
of quantum mechanics and quantum field theory:
    \be
    Z[J]:={\rm tr}\left({\cal T}
    \exp\left\{-\frac{i}{\hbar}\int_{t_1}^{t_2}(H-
    \vec J\cdot\vec x)dt\right\}\right),
    \label{Z=1}
    \ee
where ``${\rm tr}$'' stands for the trace, ${\cal T}\exp$ is the
time-ordered (chronological-ordered) exponential
\cite{weinberg,dewitt}, $t_1$ and $t_2$ are  initial and final
time labels that in field theory are taken to be $-\infty$ and
$\infty$, respectively, $\vec J\cdot\vec x=\sum_{i=1}^mJ_ix_i$ is
the source term, and $\vec x=(x_1,x_2,\cdots,x_m)$ is the
dynamical configuration variable (field).

In standard quantum mechanics the source term $\vec J\cdot\vec x$
has a specific physical meaning (most conveniently explained
within the context of Schwinger's variational principle
\cite{dewitt}) and must be a physical observable. For the case
that $\vec J$ is a real vector, which we consider here, this
entails the dynamical variables $x_i$ to be physical observables
\cite[pp 166-167]{dewitt}. In ordinary quantum mechanics this
condition is satisfied, because $x_i$ are Hermitian operators. But
in pseudo-Hermitian quantum mechanics, one must use a specific
metric operator to determine whether $x_i$ are physical
observables. In the generic case, e.g., for the cubic oscillator
(\ref{cubic-osc}), this is not the case. Therefore, the very
definition of the generating functional needs to be modified as
    \be
    Z[J]:={\rm tr}\left({\cal T}
    \exp\left\{-\frac{i}{\hbar}\int_{t_1}^{t_2}(H-
    \vec J\cdot\vec X)dt\right\}\right),
    \label{Z=}
    \ee
where $\vec X=(X_1,X_2,\cdots,X_m)$ are the pseudo-Hermitian
dynamical variables (position operators) \cite{jpa-2004b}:
    \be
    X_i:=\eta_+^{-\frac{1}{2}}x_i\eta_+^{\frac{1}{2}}.
    \label{X=}
    \ee

Without the knowledge of $\eta_+$ or the operator $Q:=-\ln\eta_+$,
one cannot construct $X_i$ and obtain the physically relevant
generating functional for the theory. The generating functional
(\ref{Z=1}) has no physical meaning unless $x_i$ commute with
$\eta_+$ (for all $i=1,2,\cdots,m$) or there is no external
sources ($\vec J=\vec 0$).

Consider the definition of ``${\rm tr}$'' that appears in
(\ref{Z=}). In the pseudo-Hermitian representation of the theory
``${\rm tr}$'' is defined by \cite{reed-simon}
    \be
    {\rm tr}(A):=\sum_n\br\psi_n,A\psi_n\kt_+=
    \sum_n\br\psi_n|\eta_+ A\psi_n\kt,
    \label{tr}
    \ee
where $\{\psi_n\}$ is an arbitrary orthonormal basis of the
physical Hilbert space ${\cal H}_{\rm phys}$ and $A$ is a
trace-class operator acting in ${\cal H}_{\rm phys}$. In view of
the fact that $\eta_+^{\frac{1}{2}}:{\cal H}_{\rm phys}\to{\cal
H}$ is a unitary operator \cite{jpa-2003}, we have
    \be
    \psi_n=\eta_+^{-\frac{1}{2}}\Psi_n
    \label{psi=psi}
    \ee
for some orthonormal basis $\{\Psi_n\}$ of ${\cal H}$. Inserting
(\ref{psi=psi}) in (\ref{tr}) and making use of (\ref{inn}) and
the Hermiticity of $\eta_+^{-\frac{1}{2}}$, we find
    \be
    {\rm tr}(A)=\sum_n\br\Psi_n|\tilde A\Psi_n\kt,
    \label{tr=tr}
    \ee
where $\tilde A:=\eta_+^{\frac{1}{2}}A\eta_+^{-\frac{1}{2}}$. The
right-hand side of (\ref{tr=tr}) is precisely the trace of the
operator $\tilde A$ that acts in ${\cal H}$. This calculation
shows that the trace of a trace-class operator $A:{\cal H}_{\rm
phys}\to{\cal H}_{\rm phys}$ is equal to the trace of its
unitary-equivalent operator, namely $\tilde A:{\cal H}\to{\cal
H}$;
    \be
    {\rm tr}(A)={\rm tr}(\tilde A).
    \label{tr=tr-A}
    \ee

Next, we apply (\ref{tr=tr-A}) for the case that $A={\cal T}
\exp\left\{-\frac{i}{\hbar}\int_{t_1}^{t_2}(H-\vec J\cdot\vec
X)dt\right\}$.  Using (\ref{Z=}) and (\ref{X=}), we have
    \be
    Z[J]={\rm tr}\left({\cal T}
    \exp\left\{-\frac{i}{\hbar}\int_{t_1}^{t_2}(h-
    \vec J\cdot\vec x)dt\right\}\right),
    \label{Z=2}
    \ee
where $h:=\eta_+^{\frac{1}{2}}H\eta_+^{-\frac{1}{2}}$ is the
equivalent Hermitian Hamiltonian \cite{jpa-2003}. It is important
to note that in the derivation of (\ref{Z=2}) we make an explicit
use of the condition that $\eta_+$ is time-independent. This is a
direct consequence of the projection axiom (measurement theory)
and unitarity \cite{plb-2007}. It allows to establish the identity
    \bea
    &&{\cal T}
    \exp\left\{-\frac{i}{\hbar}\int_{t_1}^{t_2}(h-
    \vec J\cdot\vec x)dt\right\}=\nn\\
    &&\hspace{2cm}    \eta_+^{\frac{1}{2}}
    {\cal T}
    \exp\left\{-\frac{i}{\hbar}\int_{t_1}^{t_2}(H-
    \vec J\cdot\vec X)dt\right\}\eta_+^{-\frac{1}{2}}
    \nn
    \eea
that in light of (\ref{tr=tr-A}) leads to (\ref{Z=2}).

Equation~(\ref{Z=2}) is an explicit manifestation of the physical
equivalence of Hermitian and pseudo-Hermitian representations of
the system. It is this equivalence that is used in
\cite{jones-rivers} to determine the form of the source term in
the path-integral expression for the generating functional.

Next, we consider the case that $H$ is time-independent and that
it has a real discrete spectrum and a complete set of
eigenvectors. We can always choose a set of eigenvectors $\psi_n$
of $H$ and a set of eigenvectors $\phi_n$ of $H^\dagger$ such that
$\psi_n$ and $\phi_n$ form a complete biorthonormal system, i.e.,
    \be
    \br\psi_n|\phi_m\kt=\delta_{mn},~~~~~
    \sum_n|\phi_n\kt\br\psi_n|=1,
    \label{ortho}
    \ee
and the metric operator $\eta_+$ satisfies \cite{npb-2002}
    \be
    \eta_+=\sum_n|\phi_n\kt\br\phi_n|.
    \label{eta=}
    \ee
In this case, $\{\psi_n\}$ form an orthonormal basis of ${\cal
H}_{\rm phys}$, $\eta_+$ maps $\psi_n$ to $\phi_n$, i.e.,
    \be
    \phi_n=\eta_+\psi_n,
    \label{maps}
    \ee
and in view of (\ref{Z=}), (\ref{tr}), (\ref{maps}), and the
identity $\int_{\R^m}dx_1dx_2\cdots dx_m\,|\vec x\kt\br\vec x|=1$,
we have
    \bea
    Z[\vec 0]&=&{\rm tr}\left(
    e^{-\frac{i(t_2-t_1)H}{\hbar}}\right)\nn\\
    &=&\sum_n\br\psi_n|\eta_+
    e^{-\frac{i(t_2-t_1)H}{\hbar}}\psi_n\kt\nn\\
    &=&\sum_n\int_{\R^m}dx_1dx_2\cdots dx_m\,\br\psi_n|\eta_+|
    \vec x\kt
    \br \vec x| e^{-\frac{i(t_2-t_1)H}{\hbar}}\psi_n\kt\nn\\
    &=&\int_{\R^m}dx_1dx_2\cdots dx_m\,\sum_n \br\phi_n|\vec x\kt
    \br \vec x| e^{-\frac{i(t_2-t_1)H}{\hbar}}\psi_n\kt\nn\\
    &=&\int_{\R^m}dx_1dx_2\cdots dx_m\,
    \br \vec x| e^{-\frac{i(t_2-t_1)H}{\hbar}}\sum_n
    |\psi_n\kt\br\phi_n|\vec x\kt\nn\\
    &=&\int_{\R^m}dx_1dx_2\cdots dx_m\,
    \br \vec x| e^{-\frac{i(t_2-t_1)H}{\hbar}}|\vec x\kt.
    \label{Z3}
    \eea
This calculation shows that in the absence of an external source
the generating functional does not depend on the choice of a
metric operator. It is in this sense that the metric operator
$\eta_+$ or $Q=-\ln \eta_+$ ``disappears'' from the calculations
\cite{jones-rivers}. This phenomenon is the real reason for the
metric independence of the thermodynamical quantities for the
statistical systems considered in \cite{jacobsky}. However, it
does not extend to the case where one needs to couple the system
to an external source.

As we explained above one cannot naively identify the source term
with $\vec J\cdot \vec x$, generalize (\ref{Z3}) to
    $Z[\vec J]=\int_{\R^m}dx_1dx_2\cdots dx_m\,
    \br \vec x| e^{-\frac{i(t_2-t_1)
    (H-\vec J\cdot\vec x)}{\hbar}}|\vec x\kt,$
and follow the usual root to a path-integral calculation of the
generating functional and the corresponding $n$-point functions as
done for example in \cite{bender-jpa-2006}. One must first make a
choice for the metric operator $\eta_+$, determine the observable
$\vec X$, define the generating functional according to
(\ref{Z=}), and repeat the derivation of (\ref{Z3}) with the role
of $H$ played by $H-\vec J\cdot\vec X$. This yields
    \be
    Z[\vec J]=\int_{\R^m}dx_1dx_2\cdots dx_m\,
    \br \vec x| e^{-i(\frac{t_2-t_1}{\hbar})
    (H-\vec J\cdot\vec X)}|\vec x\kt.
    \label{Z4}
    \ee
As seen from this expression the generating functional depends on
the choice of the metric operator $\eta_+$, because $\vec X=
\eta_+^{-\frac{1}{2}}\vec x \eta_+^{\frac{1}{2}}$. It is this
particular metric-dependence that makes the path-integral
calculation of the generating functional a generally formidable
task.

Unlike for the simple models considered in \cite{jones-rivers}, in
general the explicit form of an appropriate metric operator and
the pseudo-Hermitian observables that couple to sources is not
known. For example, a perturbative calculation of $h$ and $X$ for
the one-dimensional cubic oscillator (\ref{cubic-osc}) yields
\cite{jpa-2005b,jones-2005}
    \bea
    h&=&\frac{p^2}{2m}+\frac{1}{2}\mu^2x^2+
    \frac{3}{2\mu^4}\left(\frac{1}{m}\,\{x^2,p^2\}+\right.\nn\\
    &&\hspace{.5cm}\left.\mu^2x^4+
    \frac{2\hbar^2}{3m}\right)\epsilon^2+{\cal O}(\epsilon^4),
    \label{h=}\\
    X&=&x+
    \frac{2i}{m\mu^4}\left(p^2+\frac{1}{2}\,m\mu^2x^2\right)\epsilon+
    \nn\\
    &&\hspace{.5cm}\frac{1}{m\mu^6}\left(\{x,p^2\}-m\mu^2x^3\right)
    \epsilon^2+ {\cal O}(\epsilon^3),
    \label{X=new}
    \eea
where $\{\cdot,\cdot\}$ stands for the anticommutator. We can
attempt to calculate the generating functional for this system
using its Hermitian representation. This amounts to using $h-Jx$
in place of $H-JX$ in (\ref{Z4}). As shown in
\cite{bender-jpa-2006} this is already a very difficult task even
in the absence of the source term \footnote{But it makes it easy
to see that the leading order term in $Z[J]$ must be of order
$\epsilon^2$.}. If we try to use (\ref{Z4}) to compute $Z[J]$
directly, we need to insert
    \bea
    H-JX&=&\frac{p^2}{2m}+\frac{\mu^2 x^2}{2}-Jx+\nn\\
    &&\left[
    i x^3-\frac{2iJ}{m\mu^4}\left(p^2+\frac{1}{2}\,m\mu^2x^2\right)
    \right]\epsilon+\nn\\
    &&\left[-\frac{J}{m\mu^6}\left(\{x,p^2\}-m\mu^2x^3\right)\right]
    \epsilon^2+{\cal O}(\epsilon^3)
    \nn
    \eea
in (\ref{Z4}) and try to obtain a Hamiltonian path-integral
formula for $Z[J]$. Because up to terms of order $\epsilon^2$,
$H-JX$ is quadratic in $p$, $Z[J]$ admits a Lagrangian
path-integral formula that reads
    \be
    Z[J]=\int {\cal D}(x)~e^{\frac{i}{\hbar}
    \int_{t_1}^{t_2}dt~ L(x,\dot x;J)},
    \label{Z11}
    \ee
where
    \bea
    \int {\cal D}(x)&:=&\lim_{N\to\infty}\int_{-\infty}^\infty
    \cdots\int_{-\infty}^\infty
    \prod_{n=1}^N \fm(x_{n-1})dx_n,~~~~\\
    \fm(x)&:=&\left[2\pi i\hbar(t_2-t_1)N^{-1}
    g(x,J)\right]^{-\frac{1}{2}},
    \label{Z12}\\
    L(x,\dot{x};J)&:=&\frac{[\dot{x}-a(J)]^2}{2g(x,J)}-v(x,J)+
    {\cal O}(\epsilon^3),
    \label{Z13}\\
    g(x,J)&:=&m^{-1}\left[1-4i\mu^{-4}J\,\epsilon-4\mu^{-6}\epsilon^2
    \,x
    \right],
    \label{Z14}\\
    a(J)&:=&-2i\hbar m^{-1}\mu^{-6}J\,\epsilon^2,
    \label{Z15}\\
    v(x,J)&:=&\frac{1}{2}\,\mu^2x^2-Jx+(ix^3-i\mu^{-2}Jx^2)\epsilon
    \nn\\
    &&-\mu^{-4}Jx^3\,\epsilon^2.
    \label{Z16}
    \eea
The appearance of $g(x,J)$ in the expression (\ref{Z12}) for the
measure of the path integral and the kinetic part of the
Lagrangian (\ref{Z13}) is reminiscent of the perturbative
equivalence of this model and one described by a Hermitian
effective position-dependent-mass Hamiltonian
\cite{jpa-2005b,bagchi-quesne}. It is also worth mentioning that
for $J\neq 0$ the computation of the path integral (\ref{Z11}) is
as difficult as the one obtained in the Hermitian description of
the model that is based on (\ref{Z=2}).

Finally, we consider the non-Hermitian Hamiltonians of the
form
    \be
    H_\alpha=\frac{p^2}{2m}+V(x+i\alpha).
    \label{shift}
    \ee
where $V:\R\to\R$ is a real analytic function and $\alpha\in\R$.
Clearly (\ref{shift}) is obtained through an imaginary shift of
the coordinate $x$ in the Hermitian Hamiltonian $H_0$,
\cite{znojil-ahmed}. If $V$ is taken to be an even function,
(\ref{shift}) gives rise to a large class of ${\cal PT}$-symmetric
Hamiltonians:
    \be
    H=\frac{p^2}{2m}+\sum_{n=0}^\infty c_n(x+i\alpha)^{2n}=
        \frac{p^2}{2m}+\sum_{n=0}^\infty
        (-1)^nc_n(ix-\alpha)^{2n},
    \label{PT}
    \ee
where $c_n$ are real coefficients. A well-known example is
\cite{bender-prl-1998}
    \be
    H=\frac{p^2}{2m}+\frac{\mu^2x^2}{2}+i\epsilon x=
    \frac{p^2}{2m}+\frac{\mu^2}{2}(x+i\mu^{-2}\epsilon)^2+
    \frac{\epsilon^2}{2\mu^2}.
    \label{ex1}
    \ee
Another example is the following three-parameter family of ${\cal
PT}$-symmetric Hamiltonians:
    \be
    H=\frac{p^2}{2m}+
    ax^4+ibx^3+cx^2+\left[\frac{b(8ac-b^2)}{16a^2}\right](ix),
    \label{ex2}
    \ee
with $a,b,c\in\R$ and $a\neq 0$, that is obtained from (\ref{PT})
by setting $\alpha=\frac{b}{4\alpha}$, $c_2=a$,
$c_1=c+\frac{3b^2}{8a}$, $c_0=\frac{b^2(16 ac+5b^2)}{256 a^3}$,
and $c_n=0$ for $n\geq 3$.

For the case that $V$ is not an even function the Hamiltonians
(\ref{shift}) are not ${\cal PT}$-symmetric. Nevertheless (for
general $V$) they satisfy the pseudo-Hermiticity condition
\cite{p1}
    \be
    H_\alpha^\dagger=\eta_\alpha H_\alpha\eta_\alpha^{-1},
    \label{ph-alpha}
    \ee
for
    \be
    \eta_\alpha:=e^{\frac{2\alpha p}{\hbar}}.
    \label{eta-alpha}
    \ee
Because $\eta_\alpha$ is positive, $H_\alpha$ is equivalent to the
Hermitian Hamiltonian \cite{jpa-2003}
    \bea
    h&=&\eta_\alpha^{\frac{1}{2}} H_\alpha \eta_\alpha^{-\frac{1}{2}}
    \nn\\
    &=&e^{\frac{\alpha p}{\hbar}} H_\alpha
        e^{-\frac{\alpha p}{\hbar}}=
        \frac{p^2}{2m}+V(e^{\frac{\alpha p}{\hbar}}x
        e^{-\frac{\alpha p}{\hbar}}+i\alpha)
      \nn\\
      &=&\frac{p^2}{2m}+V(x)=H_0.
    \label{h=H}
    \eea
In particular, the Hamiltonians (\ref{ex1}) and (\ref{ex2}) are
respectively unitary-equivalent to the harmonic oscillator
Hamiltonian $h=\frac{p^2}{2m}+\frac{\mu^2x^2}{2}$ and the quartic
anharmonic oscillator Hamiltonian $h=\frac{p^2}{2m}+ax^4+cx^2$.

The unitary-equivalence of $H_\alpha$ and $H_0$ shows that the
generating functional $Z[J]$ for $H_\alpha$ does not dependent on
$\alpha$, i.e.,
    \bea
    Z[J]&=&\int_{-\infty}^\infty dx\,
    \br x| e^{-\frac{i(t_2-t_1)
    (H_\alpha-J X)}{\hbar}}|x\kt\nn\\
    &=&\int_{-\infty}^\infty dx\,
    \br x| e^{-\frac{i(t_2-t_1)
    (H_0-Jx)}{\hbar}}|x\kt.
    \label{z=100}
    \eea
Because of the simple form of the metric operator, we can easily
compute
    \be
    X=\eta_\alpha^{-\frac{1}{2}} x
    \eta_\alpha^{\frac{1}{2}}=x+i\alpha.
    \label{X=x3}
    \ee
Therefore, according to the first equation in (\ref{z=100}),
    \be
    Z[J]=e^{\frac{i(t_2-t_1)\alpha J}{\hbar}}\left[
    \int_{-\infty}^\infty dx\,
    \br x| e^{-\frac{i(t_2-t_1)
    (H_\alpha-J x)}{\hbar}}|x\kt\right].
    \label{Z101}
    \ee
This shows that if one uses the incorrect expression (\ref{Z=1})
for $Z[J]$ that identifies the source term with $Jx$ rather than
$JX$, one misses the factor $e^{\frac{i(t_2-t_1)\alpha J}{\hbar}}$
on the right-hand side of (\ref{Z101}). The resulting formula for
$Z[J]$ will then depend on $\alpha$ and consequently contradict
(\ref{z=100}).

In summary, we showed that one can use the standard expression for
the generating functional $Z[\vec J]$ in pseudo-Hermitian and in
particular ${\cal PT}$-symmetric quantum mechanics provided that
one modifies the source term $\vec J\cdot\vec x$ to $\vec
J\cdot\vec X$ where $\vec X$ is the pseudo-Hermitian position
operator. The metric operator $\eta_+$ that plays a central role
in the operator formulation of the theory enters in the
path-integral expression for $Z[\vec J]$ through the source term,
because $\vec X$ depends on $\eta_+$. The metric-dependence of
$\vec X$ is at the roots of the difficulty in performing the
calculations of $Z[\vec J]$ and the $n$-point functions of the
theory.


\ed